\documentclass[pdflatex,sn-mathphys-num]{sn-jnl}

\usepackage{graphicx}%
\usepackage{multirow}%
\usepackage{amsmath,amssymb,amsfonts}%
\usepackage{amsthm}%
\usepackage{mathrsfs}%
\usepackage[title]{appendix}%
\usepackage{xcolor}%
\usepackage{textcomp}%
\usepackage{manyfoot}%
\usepackage{booktabs}%
\usepackage{algorithm}%
\usepackage{algorithmicx}%
\usepackage{algpseudocode}%
\usepackage{listings}%
\usepackage{bm}
\usepackage{gensymb}
\usepackage{makecell}
\usepackage{setspace}
\usepackage{lineno}
\usepackage{geometry}

\theoremstyle{thmstyleone}%
\theoremstyle{thmstyletwo}%

\theoremstyle{thmstylethree}%

\begin{document}

\title[Fast and accurate committor estimation for kinetics simulations]{Fast and accurate committor estimation for kinetics simulations}


\author[1]{\fnm{Ru} \sur{Wang}}
\equalcont{These authors contributed equally to this work.}

\author[2]{\fnm{Xiaojun} \sur{Ji}}
\equalcont{These authors contributed equally to this work.}

\author*[1]{\fnm{Hao} \sur{Wang}}\email{wanghaosd@sdu.edu.cn}

\author[1]{\fnm{Wenjian} \sur{Liu}}

\affil[1]{\orgdiv{Qingdao Institute for Theoretical and Computational Sciences and Center for Optics Research and Engineering}, \orgname{Shandong University}, \orgaddress{\city{Qingdao}, \postcode{266237}, \state{Shandong}, \country{P. R. China}}}

\affil[2]{\orgdiv{Research Center for Mathematics and Interdisciplinary Sciences and Frontiers Science Center for Nonlinear Expectations (Ministry of Education)}, \orgname{Shandong University}, \orgaddress{\city{Qingdao}, \postcode{266237}, \state{Shandong}, \country{P. R. China}}}

\abstract{Computing long-timescale kinetics of biomolecular processes remains a major challenge for atomistic simulations. A way out is to exploit local kinetic information to construct the global stationary flux across the reaction space. The committor serves as the optimal reaction coordinate for this purpose; however, its calculation is itself highly demanding. Here, we introduce a fast and accurate algorithm for committor estimation by leveraging highly parallelizable short trajectory simulations and analogue prediction. The resulting committor is represented via a neural network ansatz and subsequently coupled with the Milestoning method to predict the mean first passage time at very low computational cost. We demonstrate the robustness and efficiency of this committor-guided Milestoning (CoM) method through examples of increasing complexity.}

\maketitle

\section{Introduction}\label{sec1}
Atomistic molecular dynamics (MD) simulation is a powerful tool for investigating thermodynamic and kinetic properties of molecular systems.
However, the timescales of many important biomolecular processes, such as protein conformational changes\cite{Shaw11} and protein-ligand unbinding\cite{HIVNNUnbind25}, exceed the reach of conventional MD simulations.
To capture such rare events, various enhanced sampling methods have been developed\cite{EnhSamp22}.
Methods designed primarily for free energy calculation typically enhance configurational sampling by adding biasing potential or forces, or by altering temperature.
They therefore struggle to recover true kinetics from the resulting distorted dynamics.
We here focus on the calculation of kinetic properties, notably the mean first passage time (MFPT), because kinetics is often more relevant than thermodynamics in living systems\cite{ModKinMile,Shaw13,ResTime16}.
Enhanced sampling methods aimed at kinetics calculation, such as Markov state model (MSM)\cite{MSM11,MSM18}, transition interface sampling (TIS)\cite{TIS05}, forward flux sampling (FFS)\cite{FFlux06,FFlux09}, weighted ensemble\cite{WE10}, adaptive multilevel splitting (AMS)\cite{AMS07,AMS19}, and Milestoning\cite{CM04,ExM15}, generally rely on unbiased trajectory simulations.
The underlying idea is to integrate local kinetic information to compute the global stationary flux through the reaction space, from which the MFPT or rate constants can be derived.

Transition path theory (TPT) provides a rigorous mathematical framework for reaction flux analysis\cite{TPT06,TPTMJ}.
The core element of TPT is the committor, defined as the probability of reaching the product state before returning to the reactant state, starting from the current configuration.
The committor is widely regarded as the optimal one-dimensional reaction coordinate, because it permits exact kinetic reconstruction\cite{CommOpt13}.
Consequently, combining the committor with many of the kinetic enhanced sampling methods mentioned above yields optimal predictions of the MFPT.
For example, iso-committor surfaces are recognized as optimal milestones in Milestoning, enabling exact MFPT calculations\cite{AssuMile08}.
Recently, the committor has also been integrated with the AMS approach.\cite{AMSComm}.

Computing the committor itself remains a major challenge for complex molecular systems.
Under overdamped Langevin dynamics, the committor satisfies the backward Kolmogorov equation (BKE)\cite{StoCal96}, which can be numerically solved only for low-dimensional model systems.
Fortunately, it can be reformulated as an equivalent variational problem\cite{CommNN18}, which, together with a neural network (NN) representation of the committor, has been successfully applied to more complex molecular systems\cite{CommNN19,CommNN26,Comm24}.
However, the variational formulation requires Boltzmann reweighting, which imposes a significant computational burden.
Moreover, the overdamped assumption may limit its accuracy when applied to molecular systems typically described by underdamped Langevin dynamics.
An alternative approach to committor calculation exploits the semigroup property: under Markovian dynamics, the committor value at a given point is related to its value after a lag time $\kappa$ via a Chapman-Kolmogorov-type equation.
Several methods follow this idea, including total squared displacement minimization\cite{Krivov15}, the Feynman-Kac variational formulation\cite{Dinn23}, and stationary flux minimization\cite{Roux22,Roux23}.
These formulations are intimately related.
A discussion of their connections can be found in the Appendix of Ref. \cite{ComminMile}.
The method we develop in Section \ref{sec_m_2} also belongs to this category, in which the semigroup formulation is built upon the analogue prediction method\cite{AnalogueP17,AnalogueP21,AnalogueP23}.
The committor calculation scheme in Ref. \cite{AMSComm} is likewise based on analogue prediction and shares a similar spirit.
However, that method relies on a single long equilibrium trajectory simulation, which is impractical for rare events in complex systems.
By contrast, our method employs highly parallelizable short trajectory simulations, whose initial configurations need not obey the Boltzmann distribution at the target condition.
This feature makes our method particularly efficient for complex systems.
The resulting committor is then represented using an NN ansatz and subsequently coupled with Milestoning for MFPT calculations, a combined approach we term committor-guided Milestoning (CoM).

In Section \ref{sec_r}, we first validate CoM on the numerically solvable Muller potential, then apply it to the classic example of alanine dipeptide in vacuum and to the folding/unfolding process of a small protein, chignolin.
Factors affecting the accuracy of CoM are discussed in Section \ref{sec_d}.
Details of the numerical implementation are provided in Section \ref{sec_m}.

\section{Results}\label{sec_r}
\subsection*{Iso-committor surfaces as optimal milestones}\label{sec_m_1}
In the Milestoning method, the phase space $\Gamma=(\mathbf{x},\mathbf{v})$ is partitioned into small compartments.
The continuous dynamics in phase space is modeled as coarse-grained stochastic transitions among the interfaces between compartments, known as milestones.
Local kinetic information, specifically the transition probabilities and mean residence time at each milestone, is collected from unbiased short trajectory simulations.
These trajectories are initiated on each milestone and propagated until they first hit an adjacent milestone.
From these simulations, the transition probability from milestone $i$ to milestone $j$ is estimated as $K_{ij}=n_{ij}/n_i$, where $n_{ij}$ is the number of trajectories starting from milestone $i$ that terminate at milestone $j$, and $n_i$ is the total number of trajectories initiated from milestone $i$.
The mean residence time for milestone $i$ is given by $t_i=\sum_{l=1}^{n_i}t_{il}/n_i$, where $t_{il}$ is the lifetime of the $l$-th trajectory initiated from milestone $i$ before it reaches any adjacent milestone.
With these two quantities, the MFPT between two predefined states (denoted by A and B) is calculated as
\begin{equation}
\tau_{A\rightarrow B}=\frac{\sum_{i\neq B}q_it_i}{q_B},
\label{tau expression}
\end{equation} 
where, without loss of generality, we assume that states A and B each contain only a single milestone. 
Generalization to states comprising multiple milestones is straightforward.
In Eq. \eqref{tau expression}, $q_i$ represents the nonequilibrium stationary flux across milestone $i$ and is obtained by solving the eigen-equation
\begin{equation}
\mathbf{q}^T\mathbf{K}'=\mathbf{q}^T,
\label{q eigeneq}
\end{equation}
where $\mathbf{K}'$ is a modified transition matrix $\mathbf{K}$ with a cyclic boundary condition imposed at state B: $K'_{Bi}=0$, for all $i\neq A$, and $K'_{BA}=1$.
This boundary condition means that whenever a trajectory reaches state B, it is immediately returned to state A.

To obtain unbiased estimates of the transition probability, mean residence time, and hence the MFPT, the initial distribution of short trajectories on the milestones must be carefully chosen.
The correct distribution is the first hitting point distribution (FHPD)\cite{AssuMile08,ExM15}.
First hitting points on a milestone $i$ are those phase space points at which an infinitely long equilibrium trajectory first arrives at milestone $i$ after having last crossed an adjacent milestone $j$ ($j\neq i$).
Unfortunately, the FHPD generally lacks an analytic expression and must be numerically approximated.
A hierarchy of approximation methods offering different trade-offs between accuracy and computational cost has been developed\cite{CM04,DiM10,LPTM23,BuM24,ExM15}.

Notably, there exists a special partitioning of phase space for which transition probabilities can be exactly calculated, even without simulations.
This is achieved by foliating the phase space with iso-committor surfaces\cite{AssuMile08}, which are therefore called optimal milestones.
Given three consecutive iso-committor surfaces, $\{C(\mathbf{x}_{i-1},\mathbf{v}_{i-1})=c_{i-1}, C(\mathbf{x}_i,\mathbf{v}_i)=c_i, C(\mathbf{x}_{i+1},\mathbf{v}_{i+1})=c_{i+1}\}$, the transition probabilities follow directly as
\begin{align}
K_{i,i+1}&=\frac{c_i-c_{i-1}}{c_{i+1}-c_{i-1}},\nonumber\\
K_{i,i-1}&=\frac{c_{i+1}-c_i}{c_{i+1}-c_{i-1}},
\label{K from C}
\end{align}
eliminating the need for short trajectory simulations.

\subsection*{Calculating the committor using analogue prediction}\label{sec_m_2}
Very often, a transition event can be effectively described by a set of collective variables (CVs), $\mathbf{\Theta}(\mathbf{x})=\{\theta_1(\mathbf{x}), \theta_2(\mathbf{x}), \cdots, \theta_s(\mathbf{x})\}$.
Under this assumption, the committor can be approximated as
\begin{equation}
C(\mathbf{x},\mathbf{v}) \approx C(\mathbf{\Theta}(\mathbf{x})).
\label{C eq}
\end{equation}
Below, we describe how to accurately estimate a committor of the form $C(\mathbf{\Theta}(\mathbf{x}))$ using the analogue prediction method.
The original analogue prediction method is a deterministic forecasting technique relying on a single long equilibrium trajectory\cite{Lorenz69,Lorenz69_2}.
Here, we introduce several key generalizations to enhance its accuracy and efficiency for rare events in complex systems.
The analogue prediction of committor (APC) algorithm proceeds as follows:
\begin{enumerate}[label=(\arabic*)]
\item The configuration space is divided into $N$ small compartments $\mathfrak{D}=\{D_1,\cdots,D_N\}$, e.g., via Voronoi tessellation, using a set of empirically chosen CVs.
These CVs need not be identical to those used to represent the committor in Eq. \eqref{C eq}.
A basic requirement is that they can distinguish the predefined states A and B.\label{C_alg_1}

\item Within each compartment $D_j\in\mathfrak{D}$, a set of initial configurations (denoted as ``$\circ$'' points) is generated using MD simulations.
During sampling, a restraining potential is applied at the compartment boundaries to prevent trajectories from escaping.
To improve sampling efficiency and ergodicity, enhanced sampling methods such as metadynamics\cite{MetaD,WellTemMetaD} or replica exchange\cite{REMD} can be employed for the generation of ``$\circ$'' points.
Importantly, the distribution of ``$\circ$'' points does not need to obey the equilibrium distribution of the target condition for which the MFPT is desired.
\label{C_alg_2}

\item For each compartment in $\mathfrak{D}$, the restraining potential is subsequently removed. 
From each sampled ``$\circ$'' point, $n$ unbiased trajectories of fixed length $\kappa$ are initiated, with velocities randomly reinitialized from the Maxwell distribution at the target temperature.
The CV values $\mathbf{\Theta}(\mathbf{x})$ of the trajectory endpoints (denoted as ``$\square$'' points) are recorded.
If a trajectory enters state A or B before time $\kappa$, it is terminated immediately upon entry.\label{C_alg_3}

\item A discrete-time Markov chain (with time step $\kappa$) is constructed among all ``$\square$'' points across all compartments $\{D_1,\cdots,D_N\}$ as follows.
For a given ``$\square$'' point, candidate destinations after time $\kappa$ are determined by its $m$ nearest ``$\circ$'' points, identified using the Euclidean distance in $\mathbf{\Theta}(\mathbf{x})$ space.
Among these $m$ ``$\circ$'' points, the $i$-th point is weighted with probability
\begin{equation}
p_i = \exp[-(d_i/\sigma)^2]/\sum_{j=1}^m\exp[-(d_j/\sigma)^2],
\end{equation}
where $d_i$ is the distance of the ``$\square$'' point to the $i$-th ``$\circ$'' point and $\sigma$ is a scaling parameter.
Subsequently, each of the $n$ terminating ``$\square$'' points associated with the $i$-th ``$\circ$'' point is assigned as a possible destination with probability $p_i/n$.
This procedure uniquely assigns transition probabilities from any ``$\square$'' point to $m\cdot n$ possible destinations, ensuring that the outgoing probabilities sum to one by construction.\label{C_alg_4}

\item Once the discrete-time Markov chain is built, the committor values for all ``$\square$'' points are computed simultaneously via iterative matrix-vector multiplication,
\begin{equation}
\mathbf{C}=\lim_{r\to\infty}(\mathbf{P}')^r\mathbf{e}_B,
\label{C pred}
\end{equation}
where $\mathbf{P}'$ is a modified transition matrix defined on the ``$\square$''-point space, with absorbing boundary conditions: $P'_{aj}=0$ for all  $a\in A$ and any $j$; $P'_{bb}=1$ and $P'_{bj}=0$ for all $b \in B$ and $j\notin B$. The vector $\mathbf{e}_B$ has entries of $1$ for all ``$\square$'' points within state B and $0$ otherwise.\label{C_alg_5} 

\item Steps \ref{C_alg_2} - \ref{C_alg_5} are iterated until the predicted committor values converge.
Convergence is monitored using the mean relative percentage error ($E_{\mathrm{MRPE}}$) of the committor between successive iterations for each compartment.
First, the relative percentage error ($E_{\mathrm{RPE}}$) for a ``$\square$'' point $\mathbf{\Theta}(\mathbf{x})$ at iteration $i$ is defined as
\begin{equation}
E_{\mathrm{RPE}}^{(i)}(\mathbf{\Theta}(\mathbf{x})) = \left|\frac{C_\mathrm{comb}^{(i)}(\mathbf{\Theta}(\mathbf{x}))-C_\mathrm{comb}^{(i-1)}(\mathbf{\Theta}(\mathbf{x}))}{C_\mathrm{comb}^{(i-1)}(\mathbf{\Theta}(\mathbf{x}))}\right|,
\label{E_RPE}
\end{equation}
where
\begin{equation}
C_\mathrm{comb}^{(i)}(\mathbf{\Theta}(\mathbf{x}))=\log_{10}(C^{(i)}(\mathbf{\Theta}(\mathbf{x}))+\epsilon)+\log_{10}(1-C^{(i)}(\mathbf{\Theta}(\mathbf{x}))+\epsilon).
\end{equation}
Here, $C^{(i)}(\mathbf{\Theta}(\mathbf{x}))$ is the committor value at iteration $i$, and $\epsilon=10^{-15}$ prevents numerical singularities.
The compartment-wise error is then $E^{(i)}_{\mathrm{MRPE}}(D_j)=\mathbb{E}_{\mathbf{\Theta}(\mathbf{x})\in D_j}\left[E^{(i)}_\mathrm{RPE}(\mathbf{\Theta}(\mathbf{x}))\right]$.
The set of compartments is updated as $\mathfrak{D}=\{D_j|E^{(i)}_{\mathrm{MRPE}}(D_j)>\alpha, j=1,\cdots,N\}$ with $\alpha=0.1$ in the present work, and Steps \ref{C_alg_2} - \ref{C_alg_5} are repeated.
For the first two iterations, $\mathfrak{D}$ includes all compartments.\label{C_alg_6}

\item Finally, a feed-forward neural network is trained on the converged committor values of the ``$\square$'' points to obtain a smooth representation of the committor.
\end{enumerate} 

\begin{figure}[h]
\centering
\includegraphics[width=1.0\textwidth]{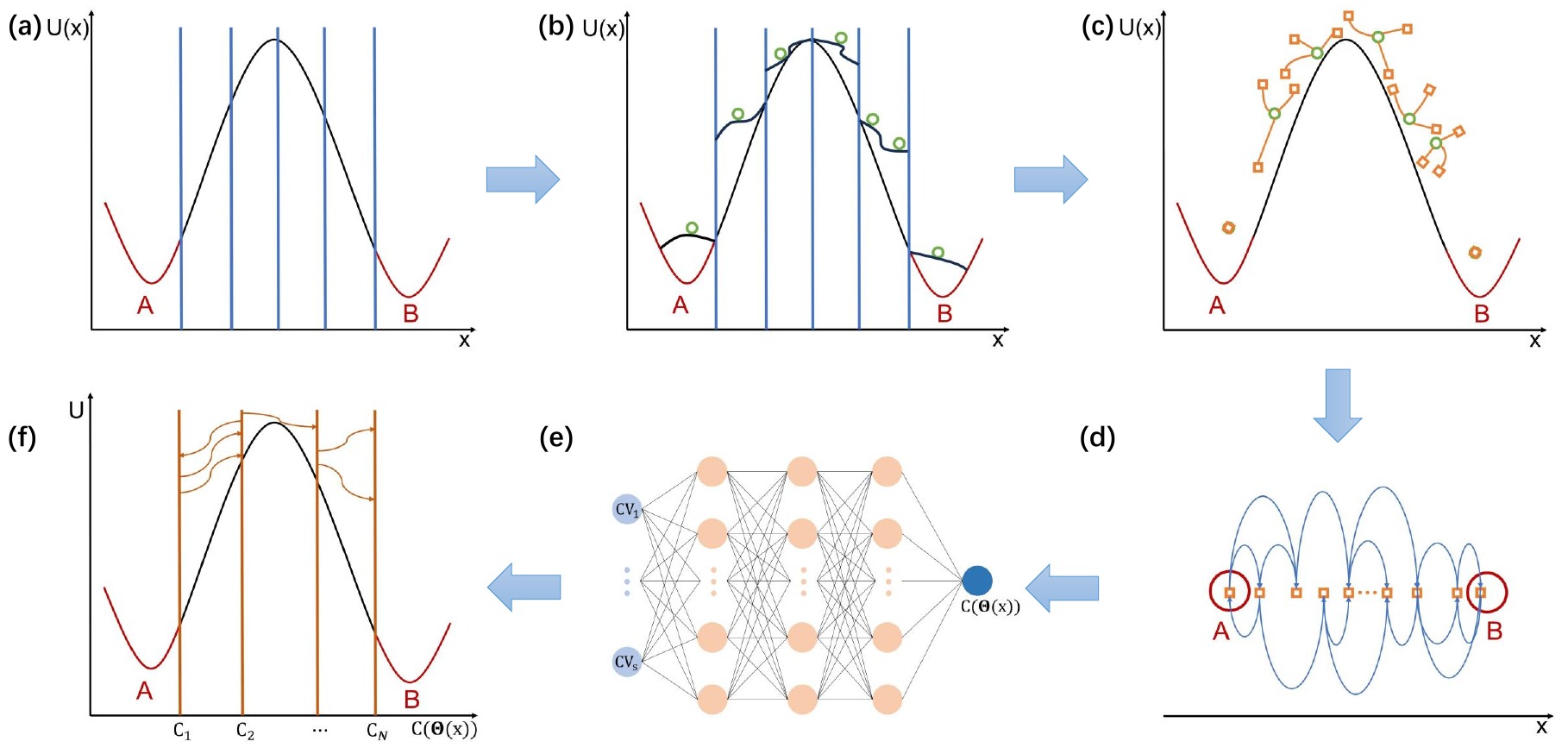}
\caption{A schematic representation of the workflow of committor-guided Milestoning (CoM). Panels (a)-(e) illustrate the flowchart of the analogue prediction of committor (APC) algorithm. (a) The configuration space is divided into small compartments. Two predefined states, A and B, are indicated in red. (b) Initial configurations (denoted as ``$\circ$'' points) are sampled within each compartment. Various enhanced sampling methods can be adopted to accelerate the sampling. (c) From each sampled ``$\circ$'' point, $n$ unbiased trajectories of fixed length $\kappa$ are initiated. Terminating points are denoted as ``$\square$'' points. (d) A discrete-time Markov chain (with time step $\kappa$) is constructed among all ``$\square$'' points based on analogue prediction. Steps (b)-(d) are repeated until convergence. (e) The converged committor prediction on the ``$\square$'' points is parameterized by a feed-forward neural network. (f) The iso-committor surfaces obtained from APC are finally coupled with Milestoning, constituting the CoM method.}\label{fig_workflow}
\end{figure}

The resulting iso-committor surfaces are then coupled with Milestoning simulations for MFPT calculations. 
The overall workflow is summarized in Fig. \ref{fig_workflow}.

\subsection*{Neural network architecture and training}\label{sec_m_3}
The committor is represented using a feed-forward neural network.
The input features are the CVs $\mathbf{\Theta}(\mathbf{x})$, optionally after a simple preprocessing transformation (see Section \ref{sec_m} for details).
The hyperbolic tangent ($\tanh$) function serves as the activation function in the hidden layers.
Since the committor is a probability bounded between $0$ and $1$, the sigmoid function is applied to the output layer.
This initial representation is denoted as $\tilde{C}_\omega(\mathbf{\Theta(\mathbf{x})})$, where the subscript $\omega$ represents the set of trainable network parameters.
To enforce the correct boundary conditions at states A and B, namely, $C(\mathbf{\Theta}(\mathbf{x}))=0$ for $\mathbf{\Theta}(\mathbf{x})\in A$ and $C(\mathbf{\Theta(\mathbf{x})})=1$ for $\mathbf{\Theta(\mathbf{x})}\in B$, we follow the approach of Ref. \cite{CommNN19} by applying a final transformation to $\tilde{C}_\omega(\mathbf{\Theta}(\mathbf{x}))$,
\begin{equation}
C_\omega(\mathbf{\Theta}(\mathbf{x}))=\bigl[1-\chi_A(\mathbf{\Theta}(\mathbf{x}))\bigr]\left[\bigl(1-\chi_B(\mathbf{\Theta}(\mathbf{x}))\bigr)\tilde{C}_\omega(\mathbf{\Theta}(\mathbf{x}))+\chi_B(\mathbf{\Theta}(\mathbf{x}))\right],
\label{NN C represent}
\end{equation}
where $\chi_{A/B}(\mathbf{\Theta}(\mathbf{x}))$ are smooth indicator functions.
The specific functional form of $\chi_{A/B}(\mathbf{\Theta}(\mathbf{x}))$ varies slightly across the test cases presented below, but all share the essential property that $\chi_{A/B}(\mathbf{\Theta}(\mathbf{x}))\approx1$ when $\mathbf{\Theta}(\mathbf{x})$ is inside state A or B, and decays rapidly to zero outside the respective state.

For training, all ``$\square$'' points and their corresponding committor values (calculated via Eq. \eqref{C pred}) are randomly split into a training set ($70\%$) and a testing set ($30\%$).
The loss function is defined as
\begin{multline}
\mathcal{L}=\frac{1}{M}\sum_{i=1}^M\left\{\Bigl[\log_{10}\bigl(C_\omega(\mathbf{\Theta}(\mathbf{x}_i))+\epsilon\bigr)-\log_{10}\bigl(C(\mathbf{\Theta}(\mathbf{x}_i))+\epsilon\bigr)\Bigr]^2\right.\\
+\left.\Bigl[\log_{10}\bigl(1-C_\omega(\mathbf{\Theta}(\mathbf{x}_i))+\epsilon\bigr)-\log_{10}\bigl(1-C(\mathbf{\Theta}(\mathbf{x}_i))+\epsilon\bigr)\Bigr]^2\right\},
\end{multline}
where $M$ is the total number of data points and $\epsilon=10^{-15}$ is a small constant to avoid numerical singularities.
The network is optimized using the mini-batch Adam optimizer with a learning rate of $10^{-3}$.
Early stopping is employed to prevent overfitting.

\subsection*{Muller potential}
The numerically solvable two-dimensional Muller potential serves as our first test case.
States A and B are defined as circles of radius $0.1$ centered at $\mathbf{z}_A=(-0.27, 1.73)$ and $\mathbf{z}_B=(0.84, 0.00)$, respectively.
To initialize the APC algorithm, the configuration space is randomly partitioned into $N=24$ compartments using Voronoi tessellation.

We first consider the overdamped dynamics, for which reference committor values can be obtained by solving the BKE with a finite difference method.
In the overdamped case, the calculated committor converges (i.e., $E_{\mathrm{MRPE}}<0.1$ for all compartments) within three iterations (see Supplementary Fig. S1).
Most sampled ``$\square$'' points are located along the minimum energy pathway.
In the final iteration, only the transition state region ($C=0.5$) exhibits a significant $E_{\mathrm{MRPE}}$, and therefore more data points are added there.
We compare a series of iso-committor surfaces ranging from state A to B generated by the APC algorithm against reference surfaces obtained from solving the BKE.
The APC-derived surfaces already show rough agreement with the reference even after the first iteration.
With successive iterations, the surfaces are progressively refined and, upon convergence, match the reference surfaces very well, including in regions where the committor is on the order of $10^{-4}$.
Noticeable errors appear mainly near the transition region but away from the transition pathway, primarily because of the scarcity of ``$\square$'' points sampled in these high-energy regions.
Since these regions are away from the transition path, they do not appreciably impact the MFPT calculation, as verified by the underdamped tests below.

\begin{figure}[h]
\centering
\includegraphics[width=1.0\textwidth]{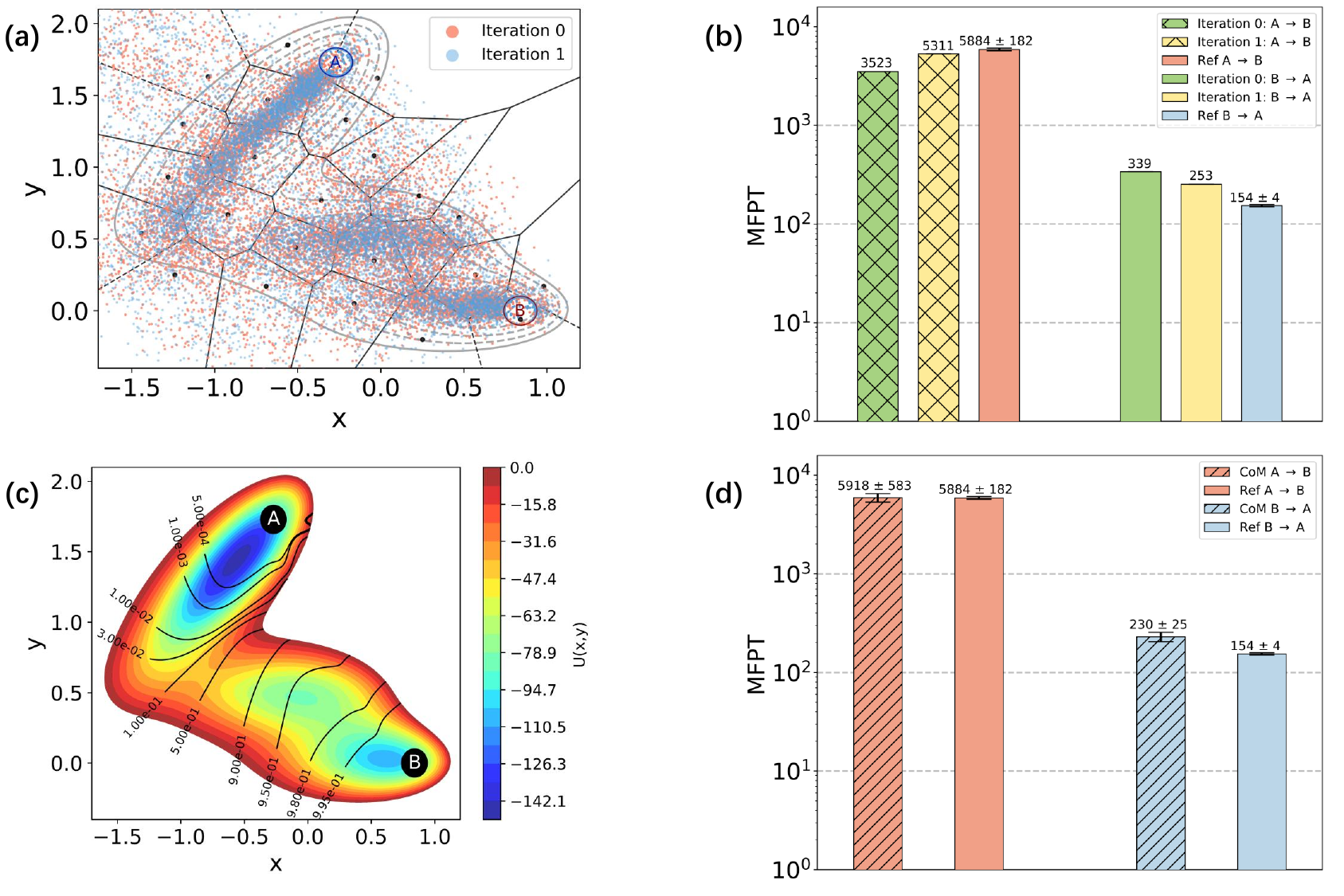}
\caption{Committor-guided Milestoning simulations for the Muller potential in the underdamped case. One of three independent experiments is shown in (a)-(c); the other two are presented in Supplementary Fig. S2. (a) The configuration space is randomly divided into 24 compartments using Voronoi tessellation. The sampled ``$\square$'' points are marked as dots, with colors indicating different iteration stages. The two states (A and B) between which the MFPT is calculated are indicated by circles. (b) The MFPT predicted after each iteration is compared with the reference value, which is obtained by running 1000 trajectories from A to B or from B to A without stopping in between with a second-order BAOAB method. (c) The iso-committor surfaces obtained from the APC algorithm for the Milestoning simulation. (d) MFPTs averaged over three independent experiments are compared with the reference value.}\label{fig_muller}
\end{figure}

As shown in Fig. \ref{fig_muller} and Supplementary Fig. S2, convergence under underdamped dynamics occurs at a similar rate. 
We conduct three independent experiments.
Typically, $2$-$3$ iterations are required to obtain a converged committor estimate, accumulating a total of $24000$-$28000$ ``$\square$'' points per experiment.
The calculated MFPTs gradually approach the reference value as iteration proceeds (Fig. \ref{fig_muller}(b)).
The reference MFPT is calculated by initializing $1000$ trajectories from the boundary of state A (or B) and simulating until they reach state B (or A) with a second-order BAOAB method\cite{BAOAB}.
Statistical errors for the calculated MFPTs are estimated from the three independent experiments.
The MFPTs averaged over these three experiments agree well with the reference value (Fig. \ref{fig_muller}(d)).
Finally, the computational cost averaged over three independent experiments is $1295.5$, which is about $22\%$ of the MFPT from state A to B.

\subsection*{Alanine dipeptide in vacuum}
Alanine dipeptide serves as another classic benchmark system for evaluating novel rare event simulation algorithms.
Its conformational transitions are often characterized by two backbone dihedral angles, $\phi$ and $\psi$.
We investigate the transition between the $\mathrm{C7}_\mathrm{eq}$ conformation (defined as a circle of radius $10^\circ$ centered at $(\phi=-75^\circ, \psi=75^\circ)$) and the $\mathrm{C7}_\mathrm{ax}$ conformation (defined as a circle of radius $10^\circ$ centered at $(\phi=80^\circ, \psi=-80^\circ)$) at $400$ K.
Here, we focus on the transition through $\phi=0^\circ$.
Accordingly, four half-harmonic restraining potentials with a force constant of $2$ kcal$\cdot$mol${^{-1}}\cdot$deg$^{-2}$ are applied at the boundaries $\phi=\pm 175^\circ$ and $\psi=\pm 175^\circ$.
To initiate the APC algorithm, the $(\phi, \psi)$ configuration space is uniformly divided into a grid of $64$ compartments, each spanning $45^\circ$ in both dimensions.

\begin{figure}[h]
\centering
\includegraphics[width=1.0\textwidth]{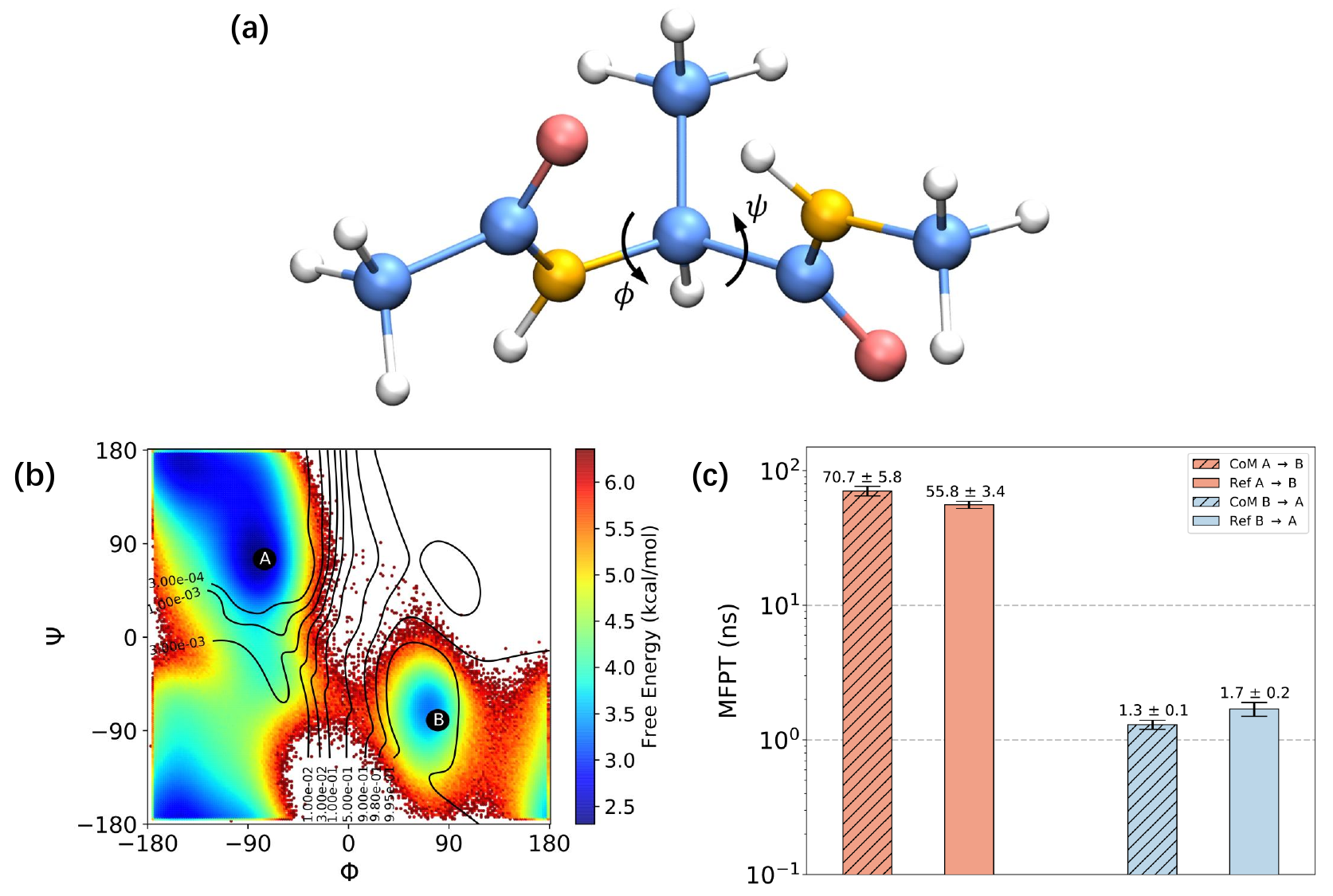}
\caption{Committor-guided Milestoning simulations for alanine dipeptide in vacuum. (a) Structure of alanine dipeptide represented in a stick-ball model: carbon in blue, oxygen in red, nitrogen in yellow, and hydrogen in white. (b) Iso-committor surfaces obtained from the APC algorithm for Milestoning simulations. The free energy landscape is obtained from 10 trajectories, each of 1 $\mu$s length. Two metastable states (A for $C7_{\mathrm{eq}}$ and B for $C7_{\mathrm{ax}}$) are indicated by circles. (c) MFPTs averaged over three independent experiments are compared with the reference value. The reference MFPT is obtained from 10 trajectories, each of 1 $\mu$s length.}\label{fig_alad}
\end{figure}

We conduct three independent experiments (see Fig. \ref{fig_alad}).
Convergence of the committor estimate typically requires $4$-$5$ iterations, accumulating $69500$-$82500$ ``$\square$'' points per experiment.
The reference MFPT is calculated from ten independent, unbiased trajectories, each of $1$ ${\mu}$s duration.
The MFPTs predicted by our method show excellent agreement with the reference values.
The computational cost averaged over three experiments is about $26$ ns, corresponding to approximately $47\%$ of the MFPT for the $C7_{\mathrm{eq}}\rightarrow C7_{\mathrm{ax}}$ transition.

\subsection*{Chignolin}
\begin{figure}[h]
\centering
\includegraphics[width=1.0\textwidth]{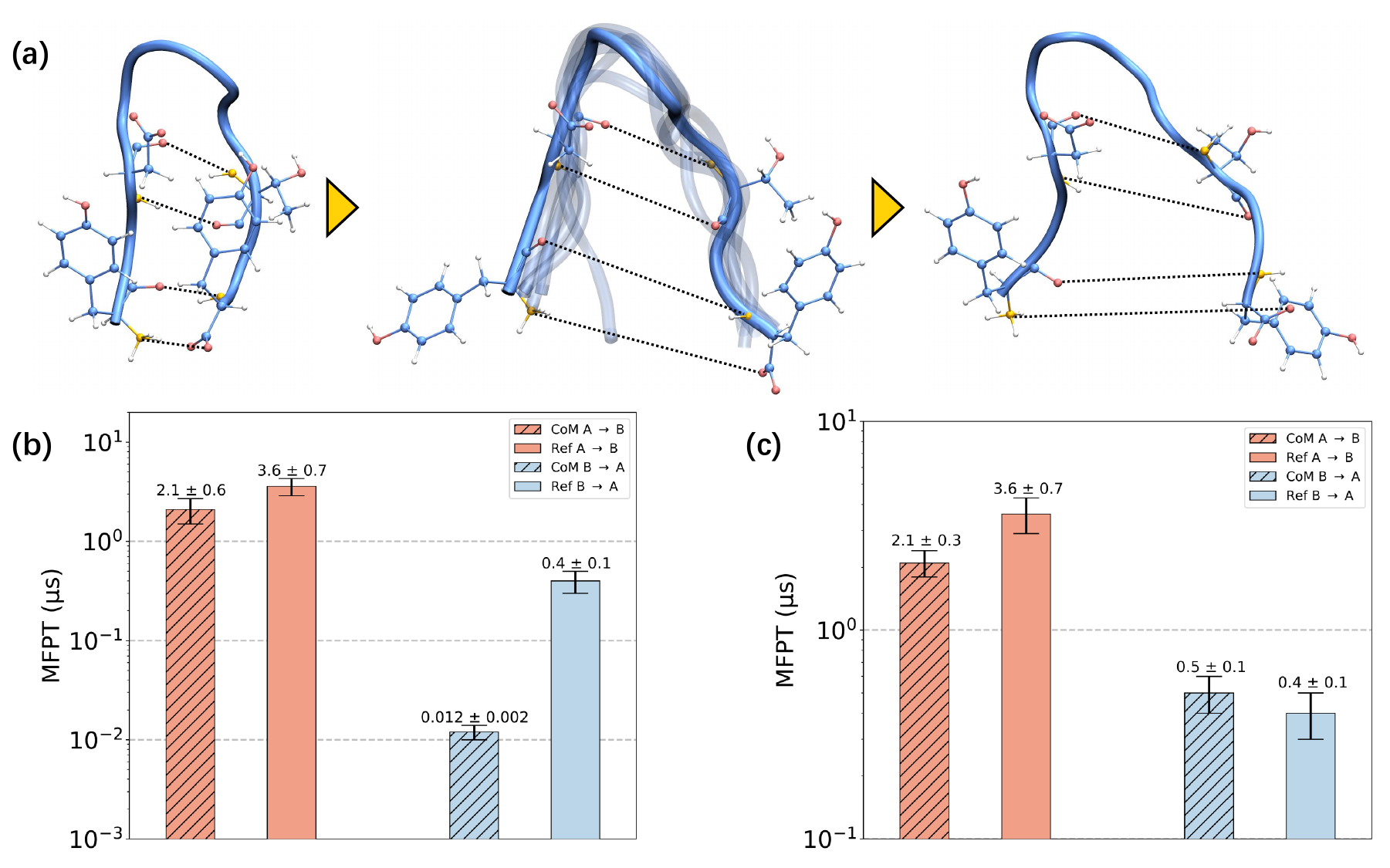}
\caption{Committor-guided Milestoning simulations for chignolin in aqueous solution. (a) Representative configurations for the folded state (denoted as state A), transition state ($C(\mathbf{\Theta}_2(\mathbf{x}))=0.5$), and unfolded state (denoted as state B). The four residues (Tyr1, Asp3, Thr8, and Tyr10) involved in the definition of folded and unfolded states are represented using a stick-ball model: carbon in blue, oxygen in red, nitrogen in yellow, and hydrogen in white. The four key distances ($d(\mathrm{Tyr1\ N}-\mathrm{Tyr10\ O})$, $d(\mathrm{Tyr1\ O}-\mathrm{Tyr10\ N})$, $d(\mathrm{Asp3\ N}-\mathrm{Thr8\ O})$, and $d(\mathrm{Asp3\ O}-\mathrm{Thr8\ N})$) used in the definition of the folded and unfolded states and in representing the committor are indicted by dashed lines. (b) MFPTs calculated with the committor represented as a function of 4 CVs ($\mathbf{\Theta}_1(\mathbf{x})$, the four distances shown in (a)). (c) MFPTs calculated with the committor represented as a function of 405 CVs ($\mathbf{\Theta}_2(\mathbf{x})$). Statistical errors in (b) and (c) are estimated from three independent experiments. The reference MFPTs are taken from Ref. \cite{Rotskoff24}.}\label{fig_chignolin}
\end{figure}

As a final and more challenging test case, we examine the folding and unfolding dynamics of chignolin, for which the MFPTs reach the microsecond scale.
We investigate the transition at $340$ K between the folded state, defined by a single CV $d_4 < 15$ \AA, and the unfolded state, defined by $d_4 > 40$ \AA.  
Here, $d_4$ is the sum of four key native contact distances: $d(\mathrm{Tyr1\ N-Tyr10\ O})$, $d(\mathrm{Tyr1\ O-Tyr10\ N})$, $d(\mathrm{Asp3\ N-Thr8\ O})$, and $d(\mathrm{Asp3\ O-Thr8\ N})$ (see Fig. \ref{fig_chignolin}).
To initiate the APC algorithm, the configuration space is partitioned into $27$ compartments along the single CV $d_4$, with the region between the folded and unfolded states uniformly divided at $1$ \AA\ intervals.
We assess the impact of CV quality on committor representation and subsequent Milestoning simulations by using two sets of CVs.

First, the committor is expressed as a function of only the four native contact distances used in the folded and unfolded state definitions, i.e., $\mathbf{\Theta}_1(\mathbf{x})=\left\{d(\mathrm{Tyr1\ N-Tyr10\ O}), d(\mathrm{Tyr1\ O-Tyr10\ N}), d(\mathrm{Asp3\ N-Thr8\ O}), d(\mathrm{Asp3\ O-Thr8\ N})\right\}$.
Across three independent experiments, convergence typically requires four iterations, accumulating $37500$-$40500$ ``$\square$'' points per experiment.
As shown in Fig. \ref{fig_chignolin}(b), the MFPT for unfolding predicted from the resulting iso-committor surfaces agrees reasonably well with the reference value ($2.1$ $\mu$s vs. $3.6$ $\mu$s).
However, the predicted MFPT for the reverse folding process shows a significant error ($0.012$ $\mu$s vs. $0.4$ $\mu$s).

We then represent the committor using a substantially expanded set of CVs, $\mathbf{\Theta}_2(\mathbf{x})$, which includes $405$ distances between the $C_\alpha$, N, and O atoms of all non-identical residue pairs.
This set is constructed in a deliberately naive manner.
A more judicious selection could likely reduce the number of CVs required. 
With this expanded representation, convergence is achieved in $3$-$4$ iterations across three experiments, accumulating $30000$-$43500$ ``$\square$'' points per experiment.
As shown in Fig. \ref{fig_chignolin}(c), the MFPTs for both folding and unfolding now agree well with the reference values.

\begin{figure}[h]
\centering
\includegraphics[width=1.0\textwidth]{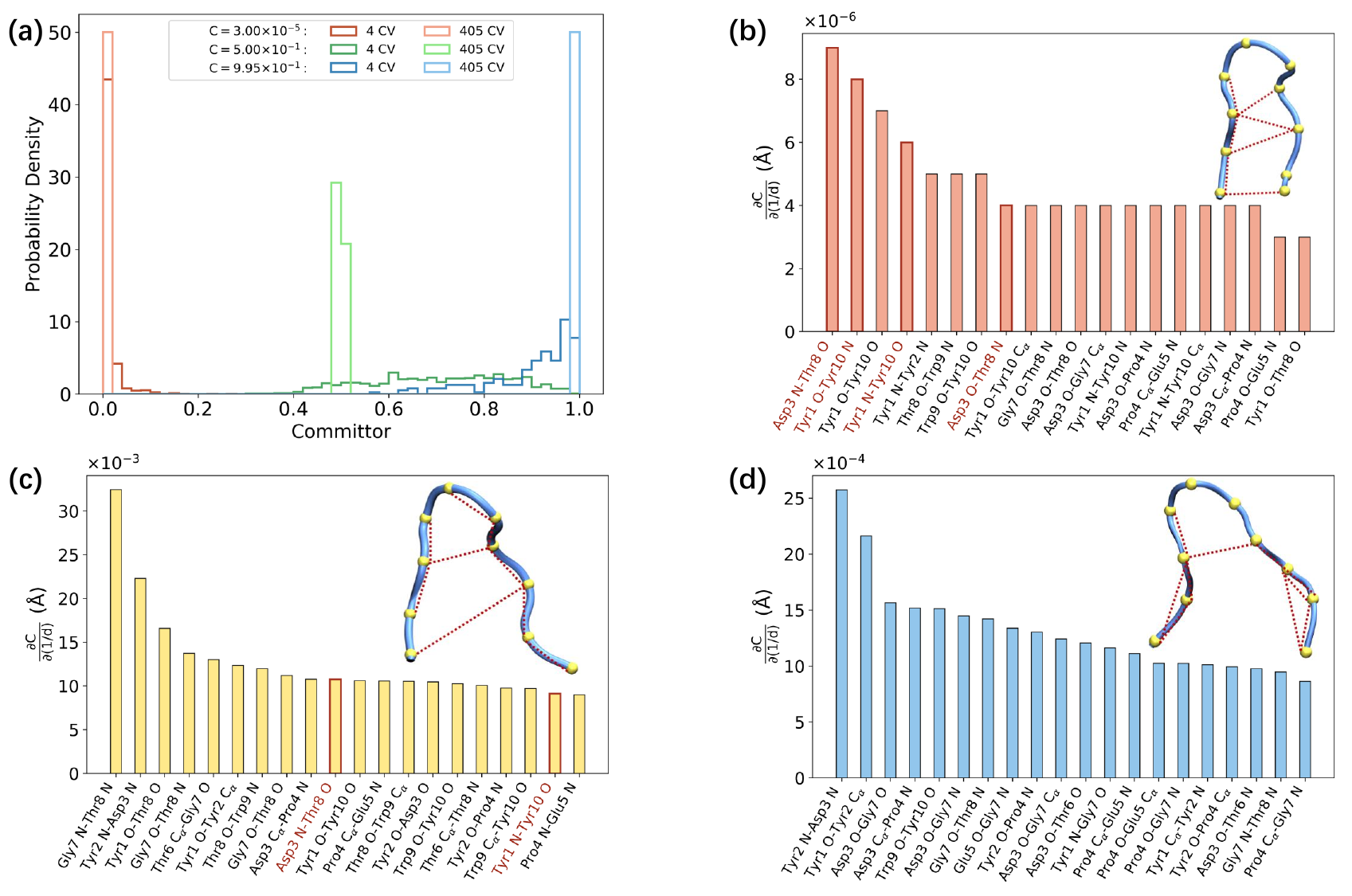}
\caption{Committor analysis for chignolin in aqueous solution. (a) Comparison of the committor distribution of ``$\square$'' points as a function of 4 CVs ($\mathbf{\Theta}_1(\mathbf{x})$) and 405 CVs ($\mathbf{\Theta}_2(\mathbf{x})$) near three iso-committor surfaces identified from $C(\mathbf{\Theta}_2(\mathbf{x}))$. (b)-(d) Sensitivity analysis of $C(\mathbf{\Theta}_2(\mathbf{x}))$ with respect to its input features. Three independent experiments are performed and show similar trends; one representative experiment is shown here, and the other two are provided in Supplementary Fig. S3. The averaged absolute gradient of the committor with respect to each input feature, $\partial C/\partial(1/d)$ for $d\in\mathbf{\Theta}_2(\mathbf{x})$, is calculated near (b) the folded state, (c) the transition state, and (d) the unfolded state. The top twenty features with the largest gradients are listed in descending order. The four CVs in $\mathbf{\Theta}_1(\mathbf{x})$ are highlighted in red. In the insets, the alpha carbon atoms of all ten residues are shown as yellow spheres, and the first five distance features that contribute the most and are identified across three independent experiments are indicated by red dashed lines.}\label{fig_chignolin_ana}
\end{figure}

The discrepancy in the MFPTs for folding obtained with the smaller CV set is attributed to inaccuracies in its iso-committor surfaces.
In Fig. \ref{fig_chignolin_ana}(a), we compare $C(\mathbf{\Theta}_1(\mathbf{x}))$ and $C(\mathbf{\Theta}_2(\mathbf{x}))$ near three iso-committor surfaces identified from $C(\mathbf{\Theta}_2(\mathbf{x}))$: one near the folded state, one in the transition state, and one near the unfolded state.
Here, the committor $C(\mathbf{\Theta}_2(\mathbf{x}))$ is used as the reference.
It can be seen that $C(\mathbf{\Theta}_1(\mathbf{x}))$ and $C(\mathbf{\Theta}_2(\mathbf{x}))$ agree well near the folded state, but differ significantly near the transition state and unfolded state, where $C(\mathbf{\Theta}_1(\mathbf{x}))$ shows a very broad distribution.
To understand this behavior, we analyze the sensitivity of the committor $C(\mathbf{\Theta}_2(\mathbf{x}))$ by computing the average gradient with respect to its $405$ input features.
As shown in Fig. \ref{fig_chignolin_ana}(b) and Supplementary Fig. S3, near the folded state (where $C(\mathbf{\Theta}_2(\mathbf{x}))$ lies in the range $[2.7\times10^{-5}, 5.7\times10^{-5}]$), at least some of the four CVs in $\mathbf{\Theta}_1(\mathbf{x})$ consistently rank among the top five in gradient magnitude, indicating their significant role in committor prediction.
Given the well-defined, conservative structure of the folded state, it is reasonable that $C(\mathbf{\Theta}_1(\mathbf{x}))$ remains accurate in this region.
However, as the system moves toward the unfolded state, the importance of these four native contacts diminishes. 
Near the transition state ($C(\mathbf{\Theta}_2(\mathbf{x}))$ in $[0.48, 0.52]$), distances between adjacent residues start to play a more dominant role (Fig. \ref{fig_chignolin_ana}(c)).
This trend continues near the unfolded state ($C(\mathbf{\Theta}_2(\mathbf{x}))$ in $[0.9935, 0.9955]$; Fig. \ref{fig_chignolin_ana}(d)).
This observation is not surprising: as the protein structure becomes increasingly flexible during unfolding, a broader set of CVs is expected to be required to adequately describe the configuration for accurate committor prediction.

The average computational cost, encompassing both committor prediction and Milestoning simulations across three independent experiments, is about $158$ ns for the $4$-CV case and $96$ ns for the $405$-CV case.
Both costs constitute less than $5\%$ of the reference unfolding MFPT.

A conventional Milestoning simulation is also performed for comparison (see Supplementary Information for details).
Thirteen milestones are defined along the single CV $d_4$.
One hundred unbiased trajectories are initiated from each milestone to estimate the transition probabilities and mean residence time.
Three independent experiments are performed.
The predicted MFPTs for unfolding and folding are $0.25\pm0.07\ \mu$s  and $5.7\pm0.7$ ns, respectively. 
The average computational cost is about $310$ ns. 

\section{Discussion}\label{sec_d}
In this work, we have developed the APC algorithm for fast and accurate estimation of the committor in complex systems, leveraging short trajectory simulations and analogue prediction.
In principle, it can be coupled with various kinetic enhanced sampling methods, such as TIS, FFS, and Milestoning.
Here, we specifically showcase the committor-guided Milestoning (CoM) approach.  
A key advantage of CoM over other Milestoning variants\cite{CM04,DiM10,ExM15,LPTM23,BuM24} is its ability to compute the local transition probability exactly from the committor, without requiring additional simulations.
This is crucial because, in MFPT calculations, the accuracy of transition probabilities is typically far more important than that of the mean residence time.
Of course, this assertion holds only when a sufficient number of milestones are placed.
In the limiting case of only two milestones (one for state A and the other for state B), the MFPT from A to B is exactly the mean residence time on milestone A. 
However, as the number of milestones increases, the importance of transition probabilities grows rapidly.
In this scenario, the nonequilibrium stationary fluxes across milestones, solved from Eq. \eqref{q eigeneq}, usually span many orders of magnitude for rare events, thereby outweighing the mean residence time in the MFPT calculation (see Eq. \eqref{tau expression}).
Consequently, in nearly all practical applications, transition probabilities exert a more significant influence on the MFPT than the mean residence time.

It should be noted that the approximate nature of the iso-committor surfaces generated by the APC algorithm does not compromise the exactness of transition probability calculation, since the latter is derived directly from the committor value defining these surfaces (Eq. \eqref{K from C}).
Approximate iso-committor surfaces are introduced only for the purpose of estimating the mean residence time on the true surfaces.
When targeting the MFPT from state A to B, particular emphasis should be placed on the accuracy of the first few iso-committor surfaces near state A, and vice versa.
This is because the stationary fluxes through these initial surfaces are orders of magnitude larger than those through subsequent ones, and according to Eq. \eqref{tau expression}, their mean residence times already largely determine the overall MFPT.

The accuracy of the APC algorithm depends on two critical factors. 
(i) Choice of the lag time $\kappa$. The lag time $\kappa$ between a ``$\circ$'' point and its corresponding ``$\square$'' points (Step \ref{C_alg_3}) needs to be long enough to ensure Markovian dynamics for the transition.
This forms the foundation for constructing the discrete-time Markov chain in Step \ref{C_alg_4}.
However, an excessively long $\kappa$ carries the risk of undersampling in high-energy regions, such as the transition state, potentially leading to a disconnected Markov chain.
Consequently, selecting $\kappa$ involves a trade-off between achieving Markovicity and maintaining adequate sampling in critical barrier regions.
The optimal $\kappa$ is intrinsically linked to the quality of the CVs chosen to represent the committor and to the intrinsic timescales of the system.
In an ideal scenario where the CVs capture all relevant slow degrees of freedom, $\kappa$ can be quite short (e.g., on the order of picoseconds, commensurate with the velocity decorrelation time in biomolecules).
Otherwise, a longer $\kappa$ is required to wash out memory effects.
Recent advances in the construction of good reaction coordinates\cite{RCReview} can improve CV quality, thereby allowing the use of shorter lag times.
The robustness of the APC and CoM methods is demonstrated by the chignolin example: a $4$-CV representation, while incomplete, yields reasonable MFPT estimates for unfolding with $\kappa=3$ ps, aided by the conservative structure of the folded state.
A comprehensive (albeit redundant) $405$-CV representation achieves accurate MFPTs for both folding and unfolding with a shorter $\kappa=1$ ps.
This multi CV set analysis can be done simultaneously in a single implementation of the APC algorithm and can be used as a validation of predicted committor and consequent MFPTs, whose computational overhead is negligible compared to the cost of atomistic simulations.
Further validation on more complex systems remains a subject for future study.
(ii) Density of sampling for analogue prediction. The analogue prediction requires a high density of ``$\circ$'' points, particularly along dominant transition pathways, to ensure that every ``$\square$'' point has sufficiently close analogues.
This requirement is primarily addressed by the initial stratification (Step \ref{C_alg_1}) combined with enhanced sampling techniques (Step \ref{C_alg_2}).
A key advantage of the APC algorithm is that the ``$\circ$'' points need not be sampled from the equilibrium distribution at the target condition, eliminating the need for Boltzmann reweighting.
Furthermore, an adaptive strategy (Step \ref{C_alg_6}) is implemented to iteratively identify regions with high committor estimation error and to augment sampling specifically in those areas.

\section{Methods}\label{sec_m}
\subsection*{Simulation protocol}\label{sec_m_4}
The overall computational workflow comprises two sequential steps. 
First, the APC algorithm is employed to estimate committor values for a set of configurations (``$\square$'' points) spanning the relevant regions of the configuration space.
At the end of the APC algorithm, an NN is trained on these computed committor values to obtain a smooth representation of the committor.
Finally, the configuration space is foliated using a series of iso-committor surfaces, on which the Milestoning method is applied to compute the MFPT.

In Milestoning, the transition probabilities between adjacent milestones are directly computed from the committor via Eq. \eqref{K from C}, eliminating the need for additional simulations.
However, estimating the mean residence time on each milestone still requires running short unbiased trajectories initiated from the milestones.
Conventionally, such initial configurations are generated using restrained sampling on the milestone hyperplane. 
Here, we adopt a simpler initialization strategy by directly collecting ``$\square$'' points on intermediate milestone.
Details of the initial configuration preparation are provided in the Supplementary Information.

All atomistic simulations are performed using NAMD 2.14\cite{NAMD}.
Key parameters for committor calculations, NN training, and Milestoning simulations are summarized in Table \ref{Table params}.

\begin{table}[h]
\caption{Key parameters for committor calculations, neural network (NN) training, and Milestoning simulations.}\label{tab2}
\begin{tabular*}{\textwidth}{@{\extracolsep\fill}lcccccc}
\toprule%
& & \multicolumn{2}{@{}c@{}}{Muller potential} & \multicolumn{1}{@{}c@{}}{Alanine dipeptide} & \multicolumn{2}{@{}c@{}}{Chignolin} \\\cmidrule{3-4}\cmidrule{5-5}\cmidrule{6-7}%
& Parameters\footnotemark[1] & Overdamped & Underdamped & - & $4$ CVs & $405$ CVs \\
\midrule
\multirow{6}{*}{Committor} & $N$  & 24 & 24 & 64 & 27 & 27 \\
                           & $N_{\circ}$ & 100 & 100 & 100 & 100 & 100 \\
                           & $n$ & 5 & 5 & 5 & 5 & 5\\
                           & $\kappa$ & $10^{-2}$ & $3\times10^{-2}$ & 300 fs & 3 ps & 1 ps\\
                           & $m$ & 10 & 10 & 20 & 50 & 50 \\
                           & $\sigma$ & 0.1 & 0.1 & 10\degree & 0.2 \AA & 4 \AA \\
\midrule
\multirow{4}{*}{NN}        & $N_{\mathrm{feature}}$ & 2 & 2 & 2 & 4 & 405 \\
                           & $N_{\mathrm{hidden}}$ & 3 & 3 & 3 & 3 & 3 \\
                           & $N_{\mathrm{neuron}}$ & 8 & 8 & 96 & 96 & 96\\
                           & $d_s$ & 0.02 & 0.02 & 2\degree & 0.2 \AA & 0.2 \AA \\
\midrule
\multirow{2}{*}{Milestoning} & $N_{\mathrm{miles}}$ & 12 & 12 & 12 & 13 & 13 \\
                             & $N_{\mathrm{unbiased}}$ & 10 & 10 & 10 & 10 & 10\\
                                           
\botrule
\end{tabular*}
\footnotetext[1]{$N$: Number of small compartments generated in Step \ref{C_alg_1} of the APC algorithm. $N_{\circ}$: Number of newly added ``$\circ$'' points per compartment per iteration (Step \ref{C_alg_2}). $n$: Number of trajectory swarms initiated from each ``$\circ$'' point to generate ``$\square$'' points (Step \ref{C_alg_3}). $\kappa$: Maximum length of each trajectory swarm (Step \ref{C_alg_3}). $m$: Number of neighboring ``$\circ$'' points used for analogue prediction (Step \ref{C_alg_4}). $\sigma$: Parameter for adjusting the relative weights of neighboring ``$\circ$'' points (Step \ref{C_alg_4}). $N_{\mathrm{feature}}$: Number of input features for the neural network. $N_{\mathrm{hidden}}$: Number of hidden layers. $N_{\mathrm{neuron}}$: Number of neurons per hidden layer. $d_s$: Parameter involved in the definition of the indicator function $\chi_{A/B}$. $N_{\mathrm{miles}}$: Number of milestones (i.e., iso-committor surfaces) used in Milestoning simulations. $N_{\mathrm{unbiased}}$: Number of unbiased short trajectories initiated from each milestone for estimating the mean residence time.}
\label{Table params}
\end{table}

\subsubsection*{Muller potential}
The energy function $U(x,y)$ is defined in Supplementary Eq. S1.
To evaluate the APC algorithm, we first consider the overdamped Langevin dynamics
\begin{equation}
\dot{\mathbf{z}}=-\frac{1}{\gamma}\nabla U(\mathbf{z})+\sqrt{\frac{2k_BT}{\gamma}}\bm{\eta},
\label{overdamped eq}
\end{equation}
where $\mathbf{z}=(x,y)^T$, $\gamma$ is the friction coefficient, $k_BT$ is the temperature, $\bm{\eta}$ is Gaussian white noise satisfying $\langle\bm{\eta}\rangle=\mathbf{0}$ and $\langle\eta_i(t)\eta_j(t')\rangle=\delta_{ij}\delta(t-t')$, with $i,j=x,y$.
All simulations use dimensionless units with $\gamma=10$.
To accelerate the sampling of ``$\circ$'' points within each compartment (Step \ref{C_alg_2} of the APC algorithm), the temperature is temporarily raised to $k_BT=20$, which is approximately $20\%$ of the energy barrier along the minimum energy path from state A to B.
During the sampling of ``$\circ$'' points, a half-harmonic restraining potential $U_b=\frac{1}{2}k_\mathrm{cont}(d_i-d_j)^2$ (applied when $d_i>d_j$ for $j\neq i$) is imposed at the boundary of the sampled compartment $D_i$, where $d_i$ ($d_j$) is the distance to the anchor point of $D_i$ ($D_j$), and the force constant $k_\mathrm{cont}=8000$.
Equation \eqref{overdamped eq} is integrated using the Euler-Maruyama scheme with a time step $\Delta t=5\times 10^{-4}$.
During each iteration, $100$ new ``$\circ$'' configurations are added to the active compartments, saved every $200$ integration steps.
From each new ``$\circ$'' point, five unbiased trajectories are propagated at the target temperature $k_BT=10$ with a time step $\Delta t=10^{-3}$. 
The maximum length of these trajectory swarms is set to $\kappa=10\Delta t=10^{-2}$.
The committor is represented by an NN using Cartesian coordinates $(x,y)$ as input features.
The indicator function in Eq. \eqref{NN C represent} is chosen as
\begin{equation}
\chi_{A/B}(\mathbf{z})=\frac{1}{2}-\frac{1}{2}\tanh\left[1000\left((\mathbf{z}-\mathbf{z}_{A/B})^2-(r_z+d_s)^2\right)\right],
\label{indicator func}
\end{equation}
where $\mathbf{z}_{A/B}$ are the centers of states A or B, $r_z$ denotes the radius of states A and B, and $d_s=0.02$ controls the smoothness near the boundaries of states A and B.

We also test the algorithm with the underdamped Langevin dynamics,
\begin{align}
\dot{\mathbf{z}}&=\mathbf{v}&\nonumber\\
m_p\dot{\mathbf{v}}&=-\nabla U(\mathbf{z})-\gamma\mathbf{v}+\sqrt{2\gamma k_BT}\bm{\eta},
\end{align}
with mass $m_p=1$ and friction $\gamma=10$.
The committor is still represented as a function of Cartesian coordinates, $C(x,y)$.
Because the dynamics of $\mathbf{z}$ is non-Markovian in the underdamped regime, the maximum swarm length is increased to $\kappa=30\Delta t=3\times 10^{-2}$.
All other settings remain identical to the overdamped case.

The final goal is to compute the MFPT between states A and B at $k_BT=10$.
For the Milestoning simulation, the configuration space is foliated using the following series of iso-committor surfaces, chosen to provide roughly uniform spacing: $[0, 5\times10^{-4}, 10^{-3}, 0.01, 0.03, 0.1, 0.5, 0.9, 0.95, 0.98, 0.995, 1]$. 
Here, the surfaces at $C=0$ and $C=1$ correspond to the boundaries of states A and B, respectively.
The integration time step of unbiased trajectories initiated from iso-committor surfaces is $10^{-4}$.

\subsubsection*{Alanine dipeptide in vacuum}
Alanine dipeptide is simulated using the CHARMM22 force field\cite{CHARMM22} with a Langevin thermostat (friction coefficient $1$ ps\textsuperscript{-1}).
To generate ``$\circ$'' points, the temperature is temporarily raised to $600$ K, using a time step of $0.5$ fs.
During the sampling of ``$\circ$'' points, a half-harmonic restraining potential with a force constant of $1$ kcal$\cdot$mol$^{-1}\cdot$deg$^{-2}$ is applied at compartment boundaries.
The swarms of trajectories initiated from ``$\circ$'' points are propagated with a $1$ fs time step at the target temperature of $400$ K. 
The committor is represented by an NN using the normalized dihedral angles $(\phi/180^\circ, \psi/180^\circ)$ as input features.
The indicator function follows the form of Eq. \eqref{indicator func} with $\mathbf{z}=(\phi, \psi)$ and $d_s=2^\circ$.
For Milestoning simulations, the iso-committor surfaces are defined at $[0, 3\times10^{-4}, 10^{-3}, 3\times10^{-3}, 0.01, 0.03, 0.1, 0.5, 0.9, 0.98, 0.995, 1]$.
The integration time step of unbiased trajectories initiated from iso-committor surfaces is $1$ fs.

\subsubsection*{Chignolin}\label{sec_m_4_chig}
The atomistic model of chignolin is constructed from the PDB structure 2RVD\cite{2RVD}.
The protein is solvated in a TIP3P water box, neutralized with $0.15$ M KCl, resulting in a system of $6927$ atoms in a periodic cubic box with a side length of $42$ \AA.
Energy minimization is performed for $10000$ steps using the conjugate gradient algorithm.
The system is then heated to $340$ K in the NVT ensemble over $1$ ns, followed by a $2$ ns NPT equilibration at $340$ K and $1$ atm using a Nose-Hoover Langevin piston\cite{NPT94,NPT95}.
Finally, the system is equilibrated in the NVT ensemble at $340$ K for an additional $1$ ns.
A Langevin thermostat with a friction coefficient of $1$ ps$^{-1}$ is used throughout.

Water molecules are constrained using the SETTLE algorithm\cite{SETTLE}, and bonds involving hydrogen atoms are fixed using the SHAKE algorithm\cite{SHAKE}.
The integration time step is 1 fs.
A real-space cutoff distance of 9.5 \AA\ is used for both electrostatic and van der Waals interactions, with long-range electrostatics handled by the particle mesh Ewald method\cite{PME}.
The CHARMM36 force field is used for the atomistic simulations\cite{CHARMM36}.

To enhance sampling of the ``$\circ$'' points within compartments, the temperature is temporarily increased to $450$ K, and configurations are saved every $2$ ps.
During the sampling of ``$\circ$'' points, a half-harmonic restraining potential with a force constant of $10$ kcal$\cdot$mol$^{-1}\cdot$\AA$^{-2}$ is applied at compartment boundaries.

The committor is represented using the inverse distances defined in either $\mathbf{\Theta}_{1}(\mathbf{x})$ or $\mathbf{\Theta}_{2}(\mathbf{x})$ as input features. 
The indicator functions for the folded (state A) and unfolded (state B) states are 
\begin{align}
\chi_{A}(\mathbf{x})&=\frac{1}{2}-\frac{1}{2}\tanh\left[1000(d_4-15+d_s)\right],\\
\chi_{B}(\mathbf{x})&=\frac{1}{2}+\frac{1}{2}\tanh\left[1000(d_4-40+d_s)\right],
\end{align}
with $d_s=0.2$ \AA.
Owing to differences in committor prediction accuracy near the unfolded state between the two CV sets, slightly different series of iso-committor surfaces are used for the Milestoning simulations.
For $\mathbf{\Theta}_1(\mathbf{x})$, the surfaces are placed at $[0, 3\times10^{-5}, 3\times10^{-4}, 3\times10^{-3}, 0.03, 0.1, 0.3, 0.5, 0.6, 0.7, 0.8, 0.9, 1]$.
For $\mathbf{\Theta}_2(\mathbf{x})$, they are $[0, 3\times10^{-5}, 3\times10^{-4}, 3\times10^{-3}, 0.03, 0.1, 0.3, 0.5, 0.7, 0.9, 0.98, 0.995, 1]$.
The integration time step of unbiased trajectories initiated from iso-committor surfaces is $1$ fs.

\section{Code and data availability}
Atomistic MD simulations were performed using NAMD 2.14, freely available at https://www.ks.uiuc.edu/Research/namd/.
Structural images were generated with the VMD software, freely available at https://www.ks.uiuc.edu/Research/vmd/.
The code and input files for the examples presented in this work are available at https://github.com/WHLab4CB/CoM.


\bibliography{CoM-bib}

\section{Acknowledgements}
This work was supported by the National Natural Science Foundation of China (No. 22403056 and No. 12401179), the Fundamental Research Funds for the Central Universities, and the Qilu Young Scholars Program of Shandong University.

\section{Author contributions}
H.W. designed the project; X.J. and H.W. developed the theoretical methods; R.W. and H.W. developed the code; R.W. conducted the simulations; R.W., X.J., and H.W. analyzed the data; and R.W., X.J., H.W., and W.L. wrote the manuscript. 

\section{Competing interests}
The authors declare no competing interest.

\end{document}